\documentclass[fleqn,10pt]{wlscirep}

\usepackage{graphicx}
\usepackage{multicol}
\usepackage{xcolor}
\usepackage{mathtools}
    \usepackage{amsmath}    
    \usepackage{amsfonts}   
    \usepackage{amssymb}    
    \usepackage{amsbsy} 
    \usepackage{amsthm} 
    \usepackage{mathptmx}   
\usepackage{rotating}
\usepackage[normalem]{ulem}
\usepackage{bm}
\usepackage{lipsum}
\usepackage{bbm}

\usepackage{bbold}

\DeclareSymbolFont{newfont}{OML}{cmm}{m}{it}
\DeclareMathSymbol{\Epsilon}{3}{newfont}{15}
\DeclareMathSymbol{\Varrho}{3}{newfont}{37}

\DeclarePairedDelimiter\abs{\lvert}{\rvert}%

\title{Zeeman-type spin splitting in non-magnetic three-dimensional compounds: Materials prediction and electrical control}

\author[1,2,*]{Carlos Mera Acosta}
\affil[1]{Center of Natural and Human Sciences, Federal University of ABC, 09210-580, Santo Andr\'e, SP, Brazil}
\affil[2]{Brazilian Nanotechnology National Laboratory, CP 6192, 13083-970, Campinas, SP, Brazil}

\author[1,2]{Adalberto Fazzio} 

\author[1]{Gustavo M. Dalpian}

\affil[*]{acosta@if.usp.br}

\begin{abstract}
\bf Despite its potential for device application, the non-magnetic Zeeman effect has only been predicted and observed in two-dimensional compounds.
We demonstrate that non-centrosymmetric three-dimensional compounds can also exhibit a Zeeman-type spin splitting, allowing the splitting control by changing the growth direction of slabs formed by these compounds.
We determine the required conditions for this effect: \textit{i}) non-centrosymmetric including polar and non-polar point groups, \textit{ii}) valence band maximum or conduction band minimum in a generic $k$-point, i.e., non-time-reversal-invariant momentum, and \textit{iii}) zero magnetic moment. Using these conditions as filters, we perform a material screening to systematically search for these systems in the aflow-ICSD database. We find 20 material candidates featuring the Zeeman-type effect. 
We also found that the spin-splitting in confined systems can be controlled by an external electric field, which in turns can induce a metal-insulator transition. We believe that the Zeeman-type effect in three-dimensional compounds can potentially be used for spin-filtering devices. 
\end{abstract}

\begin{document}

\flushbottom
\maketitle
%
%
\thispagestyle{empty}

\begin{figure}[h!]
\centering
\includegraphics[width = 0.65 \linewidth]{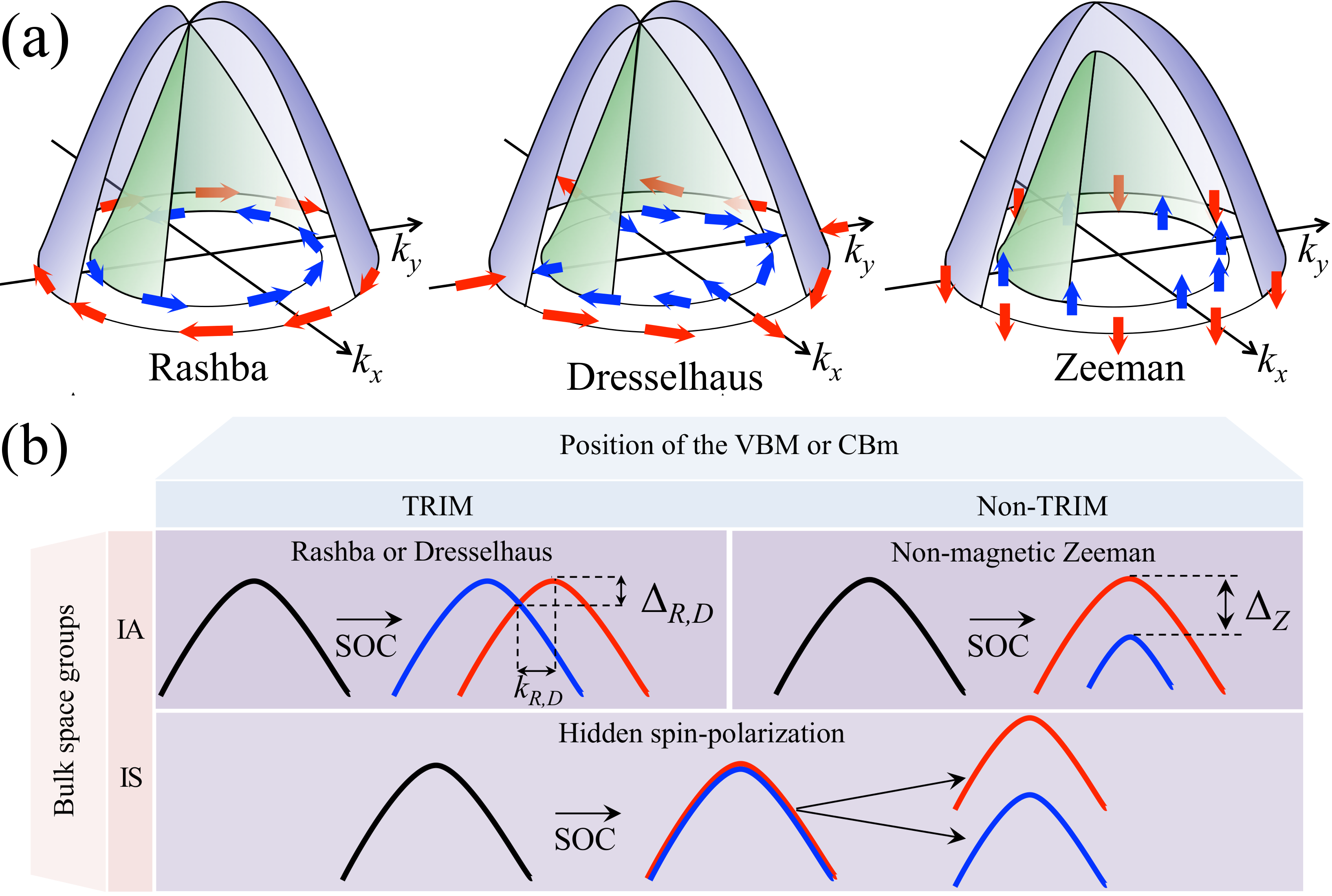}
\caption{(a) Schematic representation of the band structure and spin-texture of systems featuring the Rashba, Dresselhaus, and non-magnetic Zeeman effects. The arrows stand for the spin-polarization orientation. In the Rashba and Dresselhaus effects, the bands (green and purple) cross each other at the origin of the BZ ($\Gamma$ is a TR-invariant point). (b) Filters used in the materials prediction for each spin-polarization phenomenon. The band structure without SOC is represented in black. After considering the SOC, for IA compounds, the bands split. The up and down spins are represented in blue and red, respectively.
For compounds with IS, we have also represented each band separately to make evident the different spin components.}
\label{f:Fig0}
\end{figure}

The manipulation of inversion and time-reversal (TR) symmetries have been the cornerstone of novel phenomena allowing the generation and control of spin-polarized states in crystalline materials, the principal goal of spintronics~\cite{Ron2012,SpinCurrent2012,Manchon2015,Bercioux2015}.
The TR-symmetry breaking, which is usually induced by external magnetic fields or the intrinsic magnetic order, can lead to a separation in energy of bands with opposite spin, i.e., Zeeman spin splitting~\cite{ZEEMAN,PRESTON,PhysRevLett.113.266804}. In non-magnetic compounds, the combination of the atomic-site polarity and bulk point group results in all possible structural configurations leading to intrinsic spin-polarized states~\cite{NatAlex,HiddenOrbital,PhysRevB.96.235201}. 
For instance, in bulk inversion asymmetry (IA) materials, the spin-polarization is always accompanied by a spin splitting typically referred to as either Dresselhaus~\cite{PhysRev.100.580} or Rashba effect~\cite{RashbaSPSS60,Rashba1984} according to the spin-texture orientation (see Fig.~\ref{f:Fig0}a).
The split-bands have the opposite helical in-plane spin-texture in Rashba semiconductors and the same helicity in band inverted Rashba semiconductors~\cite{MeraAcosta2016}. 
In the Dresselhaus effect, the spin-polarization is parallel to $k$ ($\langle \vec{S}\rangle\parallel\vec{k}$) for $k_x=0$ and $k_y=0$. 
The band dispersion curves related to these effects, which are represented in Fig.~\ref{f:Fig0}b, have been characterized by spectroscopic measurements for many surfaces and interfaces~\cite{LaShell1996,Ast2007,Koroteev2004,Nitta1997}, and can be described by a simplified Hamiltonian model,
\begin{equation}
    \mathcal{H}= \mathcal{H}_{0} +
    \boldsymbol{\Omega}(\boldsymbol{k})\cdot\boldsymbol{\sigma}
    \label{2DEG}
\end{equation}
where $\mathcal{H}_{0}=\frac{\hbar^{2}k_{\parallel}^{2}}{2m^{*}}\mathbbm{1}$, $\Omega({\boldsymbol{k}})$ is the spin-orbit coupling (SOC) field, and $\sigma_{i}$ are the Pauli matrices. Here, $k_{\parallel}=k_{x}^{2}+k_{y}^{2}$, $m^{*}$ is the effective mass of electrons, and  $\mathbbm{1}$ is the 2$\times$2 unitary matrix. 
The specific form of $\boldsymbol{\Omega}(\boldsymbol{k})$
depends on the material symmetry~\cite{Vajna2012,NatComm_PST}. For instance, in a two-dimensional system with $\mathcal{C}_{3}$ point group, the Rashba and Dresselhaus fields are written as  $\boldsymbol{\Omega}_{R}=\lambda_{R}(-k_{y},k_{x},0)$ and $\boldsymbol{\Omega}_{D}=\lambda_{D}(k_{x},k_{y},0)$, respectively. The strength of the Rashba (Dresselhaus) field is given by the parameter $\lambda_{R}$ ($\lambda_{D}$). The parameter $\lambda_{R}$ is different from zero in systems featuring a non-zero electric dipole~\cite{PhysRevLett.107.156803}, which can be intrinsic or, as originally proposed by Rashba, induced by interfacing semiconductors or external electric fields. 
The Rashba effect is typically 
used for the electrical control of the spin-polarization~\cite{SpinCurrent2012,Manchon2015,Bercioux2015,CarlosMera_PRL}. For these reasons, IA materials have historically been the most promissory candidates for spintronic devices. 

Besides the Rashba and Dresselhaus effects, another kind of spin splitting in non-magnetic IA compounds, whose spin texture is similar to the one observed in the magnetic Zeeman effect, is the so-called \textit{Zeeman-type} spin splitting (See Fig.~\ref{f:Fig0}a).
Despite its potential for device application, this non-magnetic effect has only been predicted and observed in the two-dimensional WSe$_{2}$ and MoS$_{2}$~\cite{Zeeman_type,PhysRevLett.108.196802,PhysRevB.91.235204}. Unlike the Rashba and Dresselhaus splitting, the Zeeman-type splitting does not have a band crossing (See Fig.~\ref{f:Fig0}a) and has been related to the effect of an electric dipole or an external electric field~\cite{Zeeman_type}. 
This suggests that this effect can only be induced in two-dimensional systems. Indeed, it has not been explored in three-dimensional compounds.

Here, we demonstrate that IA three-dimensional compounds can also exhibit a Zeeman-type spin splitting, allowing the splitting control by changing the growth direction of slabs formed by these compounds.
For this purpose, we first 
establish the conditions for non-magnetic spin-split bands. 
Using these conditions as filters, we perform a material screening in the AFLOW-ICSD database~\cite{CURTAROLO2012227,TAYLOR2014178}, i.e., a systematic search of fabricated materials. We find 20 binary three-dimensional compounds featuring the Zeeman-type effect. 
Aside from the prediction of large Zeeman splitting in the three-dimensional version of layered systems exhibiting this effect (the VBM in MoS$_{2}$ and WS$_{2}$ have a spin splitting of 187 and 510 meV, respectively), we also find large spin splitting in compounds that are not formed by van der Waals (vdW) layered materials, such as, OsC, WN$_{2}$, and SnTe, with splittings between 330 and 490 meV. 
A representative example of these non-magnetic Zeeman materials is the SnTe compound stabilizing the Zinc-Blende (ZB) structure~\cite{Goldschmidt1927}. We confirm that in the SnTe slab, the Zeeman-type splitting depends on the growth direction, suggesting that the total electric dipole is modified by the surface induced dipole. Based on this, we then propose the electrical control of the Zeeman-type effect.
We believe that this work will open the way for the discovery of novel fundamental effects related to the spin-polarization control.

\section*{Results}
\subsection*{Design principles and materials screening} 

\begin{table*}
\caption{For each non-magnetic Zeeman-type semiconductor we show the spin splitting for the Zeeman-type effect in the CBM ($\Delta_{Z}^{\mbox{C}}$) and VBM ($\Delta_{Z}^{\mbox{V}}$) (See Fig.~\ref{f:Fig0}b), formation energy per atom ($E_{f}$), energy above the convex Hull per atom ($E_{AH}$), band gap ($E_{g}$), space group, references for the experimental realizations and previous theoretical predictions, and ICSD code. The calculated electronic properties are in agreement with the results reported in the material project~\cite{Jain2013,Jain2011a} and AFLOW-ICSD~\cite{CURTAROLO2012227,TAYLOR2014178} repositories.}
\begin{tabular}{ c c c c c c c c c c } 
\hline
Compound & $\Delta_{Z}^{\mbox{v}}$ (meV) & $\Delta_{Z}^{\mbox{c}}$ (meV) & $E_{f}$ (eV) & $E_{\mbox{AH}}$ (eV) & $E_{g}$ (eV) & Space group & Exp. & Theory  & ICSD code \\
\hline
Ge$_3$As$_4$  &  13  &  -- & 0.005 & 0.05 &	0.155 & P$\bar{4}$3m & -- & \cite{CHARIFI20091632} & 163833 \\
SnS                    &  49  &  -- & -0.495 & 0.28 & 0.166 & F$\bar{4}$3m & \cite{Badachhape1962} & \cite{Jain2013} &43409 \\
OsC                    &  332 & 340 & 0.772 & 0.772 & 0.279 & P2$_{1}$3 & -- & \cite{DuXiangP2010} & 168277 \\
Mn$_2$Ge         &  192 & 141 & 0.137 & 0.253 & 0.132 & F$\bar{4}$3m & \cite{2127} & -- & 184947 \\
RuGe                 &   97 & 101 & -0.281 & 0.013 & 0.185 & P2$_{1}$3 & \cite{Raub1962}  & -- & 637744\\
OsSi          &  346 & 202 & -0.367 & 0.012 & 0.512 & P2$_{1}$3 & \cite{Goeransson1995} & --  & 647777  \\
FeSi          &  70  &  17 & -0.511 & 0 & 0.181 & P2$_{1}$3 & \cite{Vocadlo2002,Wood1996,Pauling1948} 
& --  & 633542 \\
RuSi          &  85  &  81 & -0.647	& 0 & 0.261 & P2$_{1}$3 & \cite{Weitzer1988,Weitzer1997,Finnie1962} & -- & 85209 \\
WSi$_2$       &   -- &  38 & -0.277 & 0.066 & 0.034 & P$6_2$22  & \cite{Mattheiss1992,Heurle1980} & -- & 652549 \\
WN$_2$        & 433  & 125 & -0.521	& 0	& 1.070 & P$\bar{6}$m2 & -- & \cite{SOTO20121}  & 290433\\
Bi$_2$O$_3$   &  --  &  81 & -1.562 & 0.092 & 2.416 & R3m & -- & \cite{Matsumoto2010} & 168810\\
MoS$_2$       & 187  &  -- & -1.303	& 0.003	& 1.422 & R3m & \cite{Jellinek1960,Traill1962,Takeuchi1964} & --& 43560 \\
WS$_2$        & 510  &  -- & -1.26	& 0.004	& 1.841 & R3m & \cite{Schutte1987,Wildervanck1964} & -- & 202367 \\
Ir$_4$Ge$_5$  &  16  &  -- & -0.34	& 0 & 0.128 & P$\bar{4}$c2 & \cite{Flieher1968,Bhan1960,Panday1967} & -- & 42909 \\
Tl$_2$Te$_3$  & 156  & 197 & -0.167 & 0.024 & 0.445 & Cc & \cite{Bhan1970} &  -- & 26282 \\
GeO$_2$       &   7  &  --  & -0.533 & 1.561 &	0.985 & P3$_{1}21$ & \cite{Uno1988} & --  & 53869 \\
SnTe       & 491  &  -- & -0.034 & 0.252 & 0.240 & F$\bar{4}$3m & \cite{Goldschmidt1927} & -- & 53956 \\
Bi$_2$O$_3$   &  15  &  --  & -1.486 & 0.168 & 0.149 & P$\bar{4}$m2 &-- & \cite{Matsumoto2010} & 168808 \\
In$_2$Te$_5$  &  --  &  31 & -0.233 & 0 & 1.041 & Cc & \cite{Verkelis1974,Sutherland1976} & -- & 640615 \\
Cs$_2$Te$_3$  &  53  & 209 & -0.755 & 0 & 0.637 & Cmc2$_{1}$ & \cite{Boettcher1980,Chuntonov1982} & -- & 53245 \\
 \hline
\end{tabular}
\label{table:1}
\end{table*}

Based on the space group and chemical composition analyzes, we determine the conditions that a material should meet to exhibit the Zeeman-type effect. Design principles are typically used as filters to predict or select compounds from the materials databases~\cite{Alex_NatChem},~e.g., AFLOW~\cite{CURTAROLO2012227,TAYLOR2014178} and materials project~\cite{Jain2013,Jain2011a}. By combining the materials screening with high-throughput density functional theory calculations (See section Methods), we have an efficient approach to predict novel Zeeman-type semiconductors. 

The Zeeman-type effect is the spin discrimination as a consequence of the inversion-symmetry (IS) breaking at non-TR invariant $k$-points, rather than the effect of an intrinsic electric dipole, as we discuss below.
We then define the conditions that a material should satisfy to exhibit the Zeeman-type effect: \textit{i}) inversion-assymetry (IA) and TR symmetry (non-centrosymmetric bulk and non-magnetic moment), \textit{ii}) a total electric dipole is not needed (polar and non-polar point groups), and \textit{iii}) the VBM and CBM must take place at a non-TR-invariant $k$-point. These conditions are summarized in Fig.~\ref{f:Fig0}b.
Specifically, the Zeeman-type splitting is related to non-centrosymmetric non-magnetic materials, 
i.e., only compounds with bulk point groups $C_{n}$, $C_{nv}$, $D_{n}$ (with $n=2,3,4,$ and $6$), $S_{4}$, $D_{2h}$, $C_{3h}$, $D_{3h}$, $T$, $T_{d}$, and $O$ could feature the non-magnetic Zeeman effect. 
When the VBM or CBM take place at a TR-invariant $k$-point, the compounds are classified as a Rashba or Dresselhaus semiconductors (See Fig.~\ref{f:Fig0}b). 
Additionally, compounds in other symmetries could feature the hidden-spin-polarization, as represented in Fig.~\ref{f:Fig0}b. 

We use the previously established conditions as filters for a systematic search of non-magnetic Zeeman materials. We perform a materials screening in the AFLOW-ICSD database~\cite{CURTAROLO2012227,TAYLOR2014178}, which contains information of approximately 59000 fabricated materials. After eliminating compounds with the same formula, space group, and the number of atoms in the unit cell, we obtain 32553 entries. There are 8360 binary compounds, which in turn can be divided into 1326 IA and 7034 IS materials. Among the IA compounds, we find 587 non-magnetic gaped and 739 gapless magnetic materials. Finally, only 20 non-magnetic IA compounds exhibit either the CBM or the VBM at non-TR-invariant $k$-points. Detailed information of the predicted Zeeman-type materials is presented in Table~\ref{table:1}.

Some materials in the ICSD-database have been fabricated under specific conditions of pressure and temperature, therefore, these compounds could be above the boundary of the convex-Hull, i.e., the phase diagram representing the set of lowest possible potential energy states obtained from both single materials and mixtures of those materials. 
Remarkably, in the selected compounds, we find systems forming the boundary of the convex-Hull ($E_{AH}=0$), suggesting that they are in the most stable structural configuration, such as, WN$_{2}$ and RuSi, FeSi, Ir$_{4}$Ge$_{5}$, In$_{2}$Te$_{5}$, and Cs$_{2}$Te$_{3}$ in the space groups P$\bar{6}$m2, P2$_{1}$3, P2$_{1}$3, P$\bar{4}$c2, Cc, and Cmc2$_{1}$, respectively. We find that for RuGe, OsSi, MoS$_{2}$, WS$_{2}$, and Tl$_{2}$Te$_{3}$ the energy above the convex Hull is less than 30 meV/atom, which means that these materials could be easily synthesized.

For WS$_2$ and MoS$_{2}$ (space group $P6_{3}/mmc$), the bulk is centrosymmetric and the site point group $\mathcal{D}_{3h}$ and $\mathcal{C}_{3v}$ of the Mo and S atoms are non-centrosymmetric, which results in a hidden Dresselhaus spin-polarization~\cite{Zeeman_type,PhysRevB.91.235204}. However, in the non-centrosymmetric space group $R3m$, the VBM of these materials has a giant Zeeman-type splitting about 510 and 187 meV, as shown in Table~\ref{table:1}. 
We also find that compounds with the same formula but with different structure could exhibit different splitting values. 
For instance, Bi$_{2}$O$_{3}$ (space group $R3m$ ) has a splitting of 81 meV in the CBM, whereas the VBM of this compound in the $P\bar{4}m2$ space group has a splitting of 15 meV.
This is expected since the on-site SOC is not the unique property related to the spin splitting~\cite{PhysRevLett.107.156803}. Indeed, systems formed by atoms with a relatively weak SOC could also have large splitting values, e.g., $\Delta_{Z,V}=192$ meV in Mn$_{2}$Ge. 

\begin{figure}[h!]
\centering
\includegraphics[width = 0.95 \linewidth]{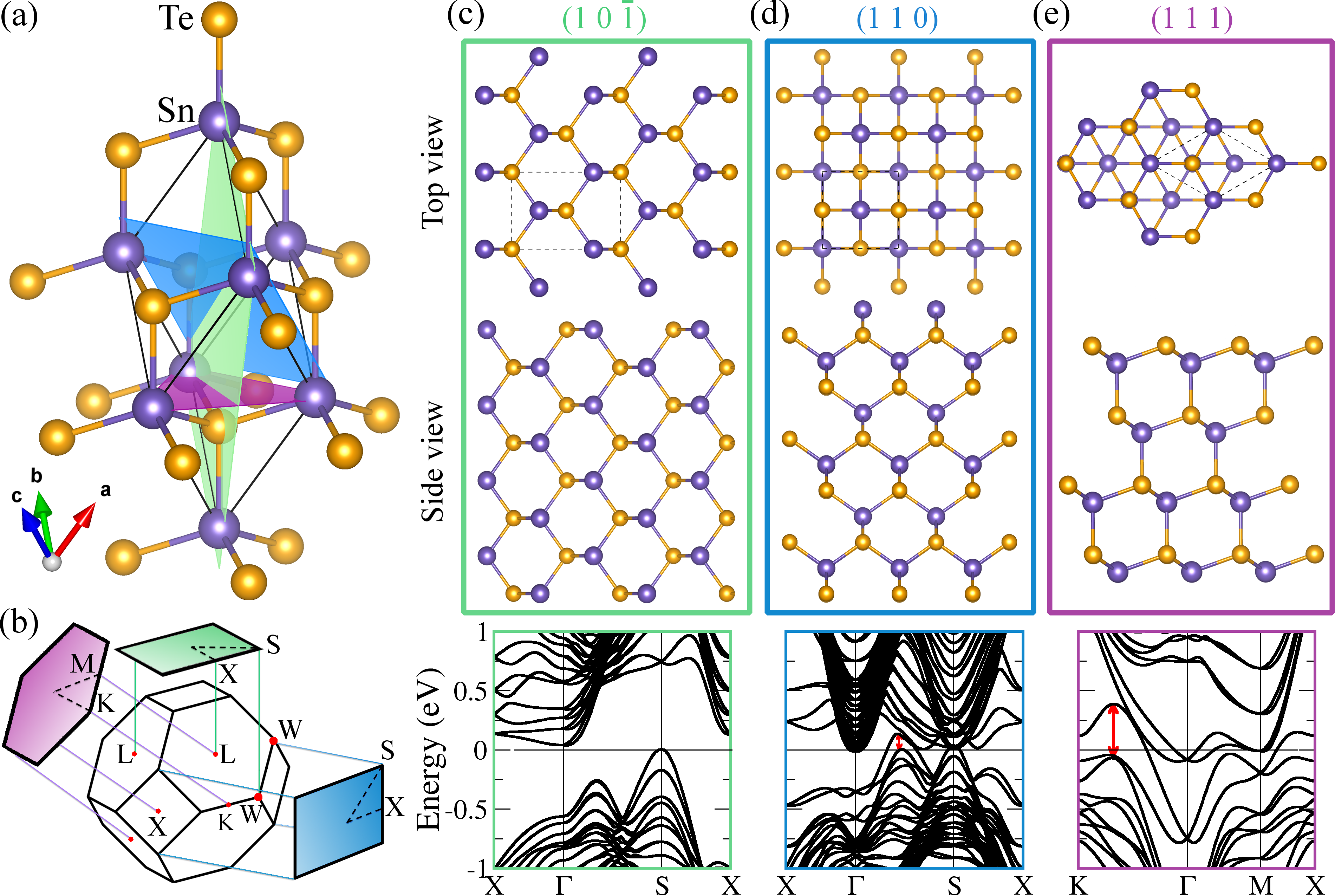}
\caption{(a) Crystal structure and (b) BZ of SnTe in the ZB structure. In (a) and (b), the crystallographic planes corresponding to slabs growth directions (10$\bar{1}$), (110), and (111) are presented by the green, blue, and purple planes, respectively. The top view and side view of the real space and band structure for the (10$\bar{1}$), (110), and (111) surfaces are shown in (c), (d), and (e), respectively. For illustrative purposes, we only show slabs formed by eight atomic layers. The red arrows stand for the spin splitting, which is zero for the (10$\bar{1}$) surface.}
\label{f:Fig1}
\end{figure}

\subsection*{Surfaces and electrical control}

Since surfaces and two-dimensional confinement affect the symmetry and the total electric dipole, we here explore the spin splitting in the surfaces of Zeeman-type semiconductors and its possible electrical control. 
For illustrative purposes, we will present here the results for the SnTe (one of the materials with the largest splitting).
We consider three growth directions corresponding to planes parallel, oblique, and perpendicular to the Sn-Te bonding along the axis normal to the surface (111) (See Fig.~\ref{f:Fig1}a), i.e., planes at the crystallographic directions (10$\bar{1}$), (110), and (111), respectively.
We find that the spin splitting change according to the growth direction of the slab. Specifically, the splitting is near zero for the plane (10$\bar{1}$) and increases as the angle between the plane and the bonding increases. Thus, the (110) and (111) planes exhibit a splitting of 50 and 491 meV  (See Fig.~\ref{f:Fig1}c-e). Consequently, the slab (10$\bar{1}$) is an insulator, whereas the (110) and (111) surfaces are metallic. 
This insulator-metal transition are a remarkable effect arising from large spin splittings in non-magnetic Zeeman semiconductors. 
If the spin splitting in the VBM is greater than the bandgap, the highest energy band can cross the Fermi energy, leading to majority spin channels, as shown for SnTe in Fig.~\ref{f:Fig1}c-e. 
This transition can be induced and controlled in a specific slab by manipulating the Zeeman splitting, which can be used for application in spintronics, e.g., spin filtering.
For instance, a perpendicular external electric field $E=E_{z}$ could modify the electric dipole induced by the surface. 
For SnTe slab along the (111) surface, an applied electric field $E=E_{z}$ decreases the spin splitting, leading to a metal-insulator transition for $E_{z}=0.12$ eV/\AA. 
By increasing the electric field, the bandgap opens again due to the change on the electric dipole orientation, resulting in an insulator-metal transition for $E_{z}=0.145$ eV/\AA.

\section*{Discussion}

\begin{figure}
\centering
\includegraphics[width = 0.55 \linewidth]{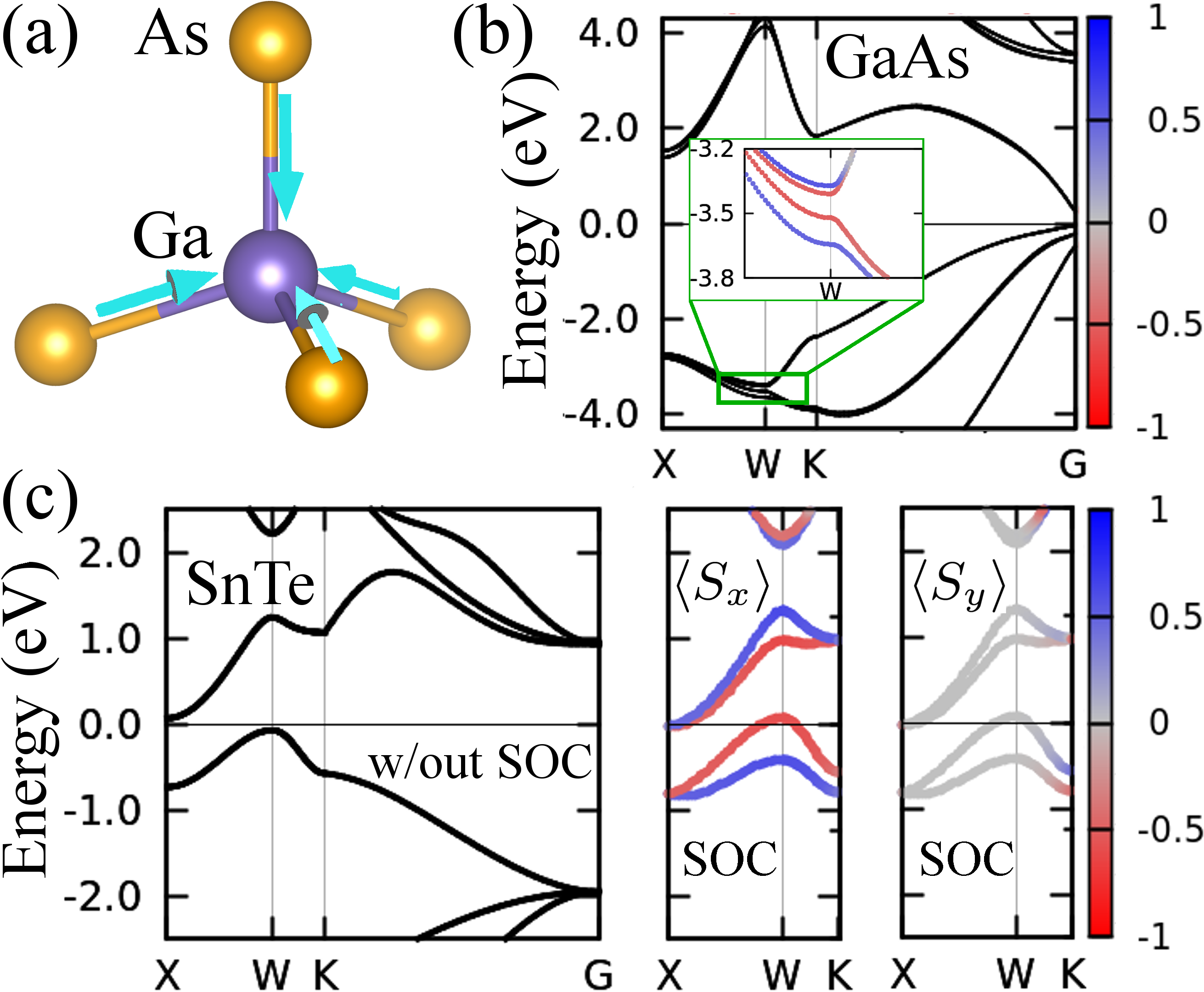}
\caption{(a) Schematic representation of the local electric dipoles formed by the Ga-As atomic interaction. Since the Ga-As bonding has the same length, the atomic-sites are non-polar ($T_{d}$ point symmetry) and the dipoles add up to zero. (b) GaAs band structure with SOC. The inset corresponds to the spin splitting at the W point. (c) Band structure without and with SOC for SnTe in the ZB structure (space group F$\bar{4}$3m). The color code stands for the orientation of the spin components. The detailed description of the spin texture is presented in the supplementary information.}
\label{f:Fig2}
\end{figure}

The Zeeman-type splitting in the so far proposed layered materials has been interpreted in terms of a non-zero intrinsic dipole~\cite{Zeeman_type}.
This interpretation, based on the common understanding of the Rashba effect, implies that compounds in which the atomic dipoles add up to zero, e.g., ZB GaAs, cannot exhibit this splitting, which is not necessarily correct, as we discuss below. 
A historical example of Dresselhaus semiconductors is the GaAs (space group F$\bar{4}$3m). In this IA compound, the dipoles add up to zero due to the tetrahedral chemical environment imposed by the crystal symmetry, as represented in Fig.~\ref{f:Fig2}a. Thus, although the Rashba terms do not contribute to the Hamiltonian describing the GaAs band structure, the spin splitting reaches high values ($\approx 120 \mbox{meV}$) at the high symmetry point W (See Fig.~\ref{f:Fig2}b), as early reported in Ref.~\cite{PhysRevLett.102.056405}. 
On the other hand, for ZB binary semiconductors, the position of the CBM and VBM can change according to the chemical composition (AB)~\cite{PhysRevB.50.2715}: GaAs and Germanium have a direct band gap at $\Gamma$, but in Silicon and GaP, the CBM is at the X point. Here, we find that the VBM can take place at the W point (highest spin splitting in GaAs) for A=Sn and B=Te, as represented in Fig.~\ref{f:Fig2}c. 
Naturally, SnTe has also a zero internal dipole, but a giant spin Zeeman-type splitting of 491 meV (See Table~\ref{table:1}). States exhibiting large spin-splitting can be brought up to the Fermi energy by changing the atomic composition, as evident from the predicted compounds. This can be a different route to find this kind of materials, which typically exhibit splittings larger than the the observed in the Rashba and Dresselhaus effects, as we demonstrate below.


In ZB semiconductors, the spin-polarized states near the $\Gamma$ point are described by the Eq.~\ref{2DEG} with  $\mathcal{H}_{0}(\boldsymbol{k})=\frac{\hbar^{2}k^{2}}{2m^{*}}\mathbbm{1}$ and 
the effective Rashba and Dresselhaus fields given by
$\boldsymbol{\Omega}_{R}(\boldsymbol{k})=\lambda_{R}(\boldsymbol{n}\times\boldsymbol{k})$ and $\boldsymbol{\Omega}_{D}(\boldsymbol{k})=\lambda_{D}\left(k_{x}(k^{2}_{y}-k^{2}_{z}),k_{y}(k^{2}_{z}-k^{2}_{x}),k_{z}(k^{2}_{x}-k^{2}_{y})\right)$, respectively. 
Here, $\boldsymbol{n}$ is a unitary vector along the direction of the electric dipole. 
Accordingly, the spin-splitting generated by these odd-in-$k$ effective magnetic fields is
\begin{equation}
\Delta(\boldsymbol{k})=\epsilon^{\uparrow}(\boldsymbol{k})-\epsilon^{\downarrow}(\boldsymbol{k})=\abs{\boldsymbol{\Omega}_{R}(\boldsymbol{k})}+\abs{\boldsymbol{\Omega}_{D}(\boldsymbol{k})},
\end{equation} 
where $\epsilon^{\uparrow}(\boldsymbol{k})$ and $\epsilon^{\downarrow}(\boldsymbol{k})$ are the eigenvalues of the Hamiltonian $\mathcal{H}$. In order to show that the Zeeman splitting is typically larger than the Rashba and Dresselhaus splitting, we consider an illustrative case: $\lambda_{D}=0$ and $\boldsymbol{n}=\hat{z}$. 
Thus, the spin-splitting becomes  $\Delta(\boldsymbol{k})=\lambda_{R}\abs{k_{\parallel}}$. In the Rashba effect, the momentum offset $k_{R}$ usually reaches small values between 10$^{-2}\mbox{\AA}^{-1}$ and 10$^{-1}\mbox{\AA}^{-1}$ (See Fig.~\ref{f:Fig0}b)~\cite{Manchon2015}. Therefore, although the parameter may be large ($\lambda_{R}\approx 1\mbox{eV\AA}$), the splitting values are always limited to some meV~\cite{Manchon2015,LaShell1996,Ast2007,Koroteev2004,Nitta1997}. This is also valid in compounds in which the atomic dipoles add up to zero, where $\lambda_{R}=0$ and $\Delta(\boldsymbol{k})=\abs{\boldsymbol{\Omega}_{D}(\boldsymbol{k})}$.
Since the non-TR-invariant high symmetry $k$-points are at the boundary of the BZ, Zeeman-type splitting are always larger than Rashba and Dresselhaus splitting. 
Additionally, due to the spin splitting position in the BZ, the Zeeman-type effect offers alternative mechanisms for spin polarization control.


Along the X-W symmetry line, the spin-texture of the SnTe band structure is dominated by $S_{x}$ spin components. This can be verified form the Hamiltonian $\mathcal{H}$ for $\lambda_{R}=0$. 
Because along such direction $k_{z}=0$ and $k^{2}_{x}\approx 0$, the Dresselhaus field can be written as $\Omega_{D}(\boldsymbol{k})\approx(k_{x}k^{2}_{y},0,0)$.
Therefore, the splitting can be controlled by changing the growth direction of slabs. For instance, when the systems are confined along the (10$\bar{1}$), we can write $\langle k_{y}\rangle=0$. Consequently, the Zeeman spin splitting at the W-point (S-point in the equivalent two-dimensional BZ, as shown in Fig.~\ref{f:Fig1}b) vanishes.  


Different from the Rashba and Dresselhaus splitting at TR-invariant $k$-points, the non-magnetic Zeeman effect does not allow the spin-polarization electrical control, but the spin-filtering effect mediated by the electrical control of the splitting size. Specifically, an external electric field does not change the spin-polarization, but the splitting size.
The electric field can then bring states with a specific spin to the Fermi energy. 
This control mechanism is different from the recently reported in the magnetic-Zeeman splitting~\cite{PhysRevLett.121.077701}.  
As previously discussed, the electrical control of the spin splitting can also be achieved in these materials according to the growth direction of the slab.

In conclusion, we demonstrate that IA three-dimensional non-magnetic compounds can exhibit a Zeeman-type spin splitting, providing the possibility to grow slabs in which the electrical dipole is perpendicular to the surface and hence, allowing the electrical control of the spin splitting.
The required conditions for this effect are: valence band maximum or conduction band minimum in a non-time-reversal-invariant k-point, inversion asymmetry, and zero magnetic moment. Using these conditions as filters, we perform a material screening and high-throughput ab-initio calculations to systematically search for these materials in the aflow-ICSD database. We find 20 candidates featuring this splitting. Our calculated spin splittings can be as large as 433, 510, and 491meV for the compounds WN$_2$ (P6m1), WS$_2$ (R3m), and SnTe (F43m), respectively. We also demonstrate that the spin splitting in slabs of these compounds depends on the growth direction and can be controlled by an external electric field. 
We believe that this work will open the way for the discovery of novel fundamental effect related to the spin-polarization control.

\section*{Methods}

The \textit{ab initio} calculations were performed within the density functional theory (DFT) using Perdew-Burke-Ernzenhof generalized gradient approximation (PBE)~\cite{PhysRevLett.77.3865} exchange-correlation functional and the Hubbard on-site term~\cite{PhysRevB.52.R5467,PhysRevB.57.1505}  as implemented in the Vienna Ab-initio Simulation Package (VASP)~\cite{PhysRevB.54.11169,PhysRevB.59.1758}. 
All the specific settings of the calculations without spin-orbit coupling (e.g. cutoff energies, k-point sampling, effective U parameters, atomic configurations) were the same as to those used on the AFLOW database~
\cite{CURTAROLO2012227,TAYLOR2014178}.
We have then included the spin-orbit interaction keeping the zero magnetic moment.

\bibliography{Ref}

\begin{thebibliography}{10}
\expandafter\ifx\csname url\endcsname\relax
  \def\url#1{\texttt{#1}}\fi
\expandafter\ifx\csname urlprefix\endcsname\relax\def\urlprefix{URL }\fi
\expandafter\ifx\csname doiprefix\endcsname\relax\def\doiprefix{DOI }\fi
\providecommand{\bibinfo}[2]{#2}
\providecommand{\eprint}[2][]{\url{#2}}

\bibitem{Ron2012}
\bibinfo{author}{Jansen, R.}
\newblock \bibinfo{journal}{\bibinfo{title}{Silicon spintronics}}.
\newblock {\emph{\JournalTitle{Nat Mater}}} \textbf{\bibinfo{volume}{11}},
  \bibinfo{pages}{400--408} (\bibinfo{year}{2012}).
\newblock \urlprefix\url{http://dx.doi.org/10.1038/nmat3293}.
\newblock \doiprefix 10.1038/nmat3293.

\bibitem{SpinCurrent2012}
\bibinfo{author}{Maekawa, S.}, \bibinfo{author}{Valenzuela, S.~O.},
  \bibinfo{author}{Saitoh, E.} \& \bibinfo{author}{Kimura, T.}
\newblock \emph{\bibinfo{title}{Spin Current}} (\bibinfo{publisher}{Series on
  Semiconductor Science and Technology 17}, \bibinfo{address}{Oxford University
  Pres}, \bibinfo{year}{2012}).

\bibitem{Manchon2015}
\bibinfo{author}{Manchon, A.}, \bibinfo{author}{Koo, H.~C.},
  \bibinfo{author}{Nitta, J.}, \bibinfo{author}{Frolov, S.~M.} \&
  \bibinfo{author}{Duine, R.~A.}
\newblock \bibinfo{journal}{\bibinfo{title}{New perspectives for rashba
  spin-orbit coupling}}.
\newblock {\emph{\JournalTitle{Nat Mater}}} \textbf{\bibinfo{volume}{14}},
  \bibinfo{pages}{871--882} (\bibinfo{year}{2015}).
\newblock \urlprefix\url{http://dx.doi.org/10.1038/nmat4360}.

\bibitem{Bercioux2015}
\bibinfo{author}{Bercioux, D.} \& \bibinfo{author}{Lucignano, P.}
\newblock \bibinfo{journal}{\bibinfo{title}{Quantum transport in rashba
  spin–orbit materials: a review}}.
\newblock {\emph{\JournalTitle{Reports on Progress in Physics}}}
  \textbf{\bibinfo{volume}{78}}, \bibinfo{pages}{106001}
  (\bibinfo{year}{2015}).
\newblock \urlprefix\url{http://stacks.iop.org/0034-4885/78/i=10/a=106001}.

\bibitem{ZEEMAN}
\bibinfo{author}{Zeeman, P.}
\newblock \bibinfo{journal}{\bibinfo{title}{The effect of magnetisation on the
  nature of light emitted by a substance}}.
\newblock {\emph{\JournalTitle{Nature}}} \textbf{\bibinfo{volume}{55}},
  \bibinfo{pages}{347 EP --} (\bibinfo{year}{1897}).
\newblock \urlprefix\url{http://dx.doi.org/10.1038/055347a0}.

\bibitem{PRESTON}
\bibinfo{author}{Preston, T.}
\newblock \bibinfo{journal}{\bibinfo{title}{Radiation phenomena in the magnetic
  field}}.
\newblock {\emph{\JournalTitle{Nature}}} \textbf{\bibinfo{volume}{59}},
  \bibinfo{pages}{224 EP --} (\bibinfo{year}{1899}).
\newblock \urlprefix\url{http://dx.doi.org/10.1038/059224c0}.

\bibitem{PhysRevLett.113.266804}
\bibinfo{author}{Li, Y.} \emph{et~al.}
\newblock \bibinfo{journal}{\bibinfo{title}{Valley splitting and polarization
  by the zeeman effect in monolayer ${\mathrm{mose}}_{2}$}}.
\newblock {\emph{\JournalTitle{Phys. Rev. Lett.}}}
  \textbf{\bibinfo{volume}{113}}, \bibinfo{pages}{266804}
  (\bibinfo{year}{2014}).
\newblock
  \urlprefix\url{https://link.aps.org/doi/10.1103/PhysRevLett.113.266804}.
\newblock \doiprefix 10.1103/PhysRevLett.113.266804.

\bibitem{NatAlex}
\bibinfo{author}{Zhang, X.}, \bibinfo{author}{Liu, Q.}, \bibinfo{author}{Luo,
  J.-W.}, \bibinfo{author}{Freeman, A.~J.} \& \bibinfo{author}{Zunger, A.}
\newblock \bibinfo{journal}{\bibinfo{title}{Hidden spin polarization in
  inversion-symmetric bulk crystals}}.
\newblock {\emph{\JournalTitle{Nature Physics}}} \textbf{\bibinfo{volume}{10}},
  \bibinfo{pages}{387 EP --} (\bibinfo{year}{2014}).
\newblock \urlprefix\url{http://dx.doi.org/10.1038/nphys2933}.

\bibitem{HiddenOrbital}
\bibinfo{author}{Ryoo, J.~H.} \& \bibinfo{author}{Park, C.-H.}
\newblock \bibinfo{journal}{\bibinfo{title}{Hidden orbital polarization in
  diamond, silicon, germanium, gallium arsenide and layered materials}}.
\newblock {\emph{\JournalTitle{Npg Asia Materials}}}
  \textbf{\bibinfo{volume}{9}}, \bibinfo{pages}{e382 EP --}
  (\bibinfo{year}{2017}).
\newblock \urlprefix\url{http://dx.doi.org/10.1038/am.2017.67}.

\bibitem{PhysRevB.96.235201}
\bibinfo{author}{Ram\'{\i}rez-Ruiz, J.}, \bibinfo{author}{Boutin, S.} \&
  \bibinfo{author}{Garate, I.}
\newblock \bibinfo{journal}{\bibinfo{title}{Nmr in an electric field: A bulk
  probe of the hidden spin and orbital polarizations}}.
\newblock {\emph{\JournalTitle{Phys. Rev. B}}} \textbf{\bibinfo{volume}{96}},
  \bibinfo{pages}{235201} (\bibinfo{year}{2017}).
\newblock \urlprefix\url{https://link.aps.org/doi/10.1103/PhysRevB.96.235201}.
\newblock \doiprefix 10.1103/PhysRevB.96.235201.

\bibitem{PhysRev.100.580}
\bibinfo{author}{Dresselhaus, G.}
\newblock \bibinfo{journal}{\bibinfo{title}{Spin-orbit coupling effects in zinc
  blende structures}}.
\newblock {\emph{\JournalTitle{Phys. Rev.}}} \textbf{\bibinfo{volume}{100}},
  \bibinfo{pages}{580--586} (\bibinfo{year}{1955}).
\newblock \urlprefix\url{https://link.aps.org/doi/10.1103/PhysRev.100.580}.
\newblock \doiprefix 10.1103/PhysRev.100.580.

\bibitem{RashbaSPSS60}
\bibinfo{author}{Rashba, E.~I.}
\newblock \bibinfo{journal}{\bibinfo{title}{{Properties of semiconductors with
  an extremum loop. 1. Cyclotron and combinational resonance in a magnetic
  field perpendicular to the plane of the loop}}}.
\newblock {\emph{\JournalTitle{Sov. Phys. Solid State}}}
  \textbf{\bibinfo{volume}{2}}, \bibinfo{pages}{1224--1238}
  (\bibinfo{year}{1960}).

\bibitem{Rashba1984}
\bibinfo{author}{Bychkov, Y.~A.} \& \bibinfo{author}{Rashba, E.~I.}
\newblock \bibinfo{journal}{\bibinfo{title}{Properties of a 2d electron gas
  with lifted spectral degeneracy}}.
\newblock {\emph{\JournalTitle{JETP Lett}}} \textbf{\bibinfo{volume}{39}},
  \bibinfo{pages}{78} (\bibinfo{year}{1984}).
\newblock
  \urlprefix\url{http://www.jetpletters.ac.ru/ps/1264/article_19121.shtml}.

\bibitem{MeraAcosta2016}
\bibinfo{author}{Mera~Acosta, C.}, \bibinfo{author}{Babilonia, O.},
  \bibinfo{author}{Abdalla, L.} \& \bibinfo{author}{Fazzio, A.}
\newblock \bibinfo{journal}{\bibinfo{title}{Unconventional spin texture in a
  noncentrosymmetric quantum spin hall insulator}}.
\newblock {\emph{\JournalTitle{Phys. Rev. B}}} \textbf{\bibinfo{volume}{94}},
  \bibinfo{pages}{041302} (\bibinfo{year}{2016}).
\newblock \urlprefix\url{https://link.aps.org/doi/10.1103/PhysRevB.94.041302}.
\newblock \doiprefix 10.1103/PhysRevB.94.041302.

\bibitem{LaShell1996}
\bibinfo{author}{LaShell, S.}, \bibinfo{author}{McDougall, B.~A.} \&
  \bibinfo{author}{Jensen, E.}
\newblock \bibinfo{journal}{\bibinfo{title}{Spin splitting of an au(111)
  surface state band observed with angle resolved photoelectron spectroscopy}}.
\newblock {\emph{\JournalTitle{Phys. Rev. Lett.}}}
  \textbf{\bibinfo{volume}{77}}, \bibinfo{pages}{3419--3422}
  (\bibinfo{year}{1996}).
\newblock \urlprefix\url{http://link.aps.org/doi/10.1103/PhysRevLett.77.3419}.
\newblock \doiprefix 10.1103/PhysRevLett.77.3419.

\bibitem{Ast2007}
\bibinfo{author}{Ast, C.~R.} \emph{et~al.}
\newblock \bibinfo{journal}{\bibinfo{title}{Giant spin splitting through
  surface alloying}}.
\newblock {\emph{\JournalTitle{Phys. Rev. Lett.}}}
  \textbf{\bibinfo{volume}{98}}, \bibinfo{pages}{186807}
  (\bibinfo{year}{2007}).
\newblock
  \urlprefix\url{http://link.aps.org/doi/10.1103/PhysRevLett.98.186807}.
\newblock \doiprefix 10.1103/PhysRevLett.98.186807.

\bibitem{Koroteev2004}
\bibinfo{author}{Koroteev, Y.~M.} \emph{et~al.}
\newblock \bibinfo{journal}{\bibinfo{title}{Strong spin-orbit splitting on bi
  surfaces}}.
\newblock {\emph{\JournalTitle{Phys. Rev. Lett.}}}
  \textbf{\bibinfo{volume}{93}}, \bibinfo{pages}{046403}
  (\bibinfo{year}{2004}).
\newblock
  \urlprefix\url{http://link.aps.org/doi/10.1103/PhysRevLett.93.046403}.
\newblock \doiprefix 10.1103/PhysRevLett.93.046403.

\bibitem{Nitta1997}
\bibinfo{author}{Nitta, J.}, \bibinfo{author}{Akazaki, T.},
  \bibinfo{author}{Takayanagi, H.} \& \bibinfo{author}{Enoki, T.}
\newblock \bibinfo{journal}{\bibinfo{title}{Gate control of spin-orbit
  interaction in an inverted
  i${\mathrm{n}}_{0.53}$g${\mathrm{a}}_{0.47}$as/i${\mathrm{n}}_{0.52}$a${\mathrm{l}}_{0.48}$as
  heterostructure}}.
\newblock {\emph{\JournalTitle{Phys. Rev. Lett.}}}
  \textbf{\bibinfo{volume}{78}}, \bibinfo{pages}{1335--1338}
  (\bibinfo{year}{1997}).
\newblock \urlprefix\url{http://link.aps.org/doi/10.1103/PhysRevLett.78.1335}.
\newblock \doiprefix 10.1103/PhysRevLett.78.1335.

\bibitem{Vajna2012}
\bibinfo{author}{Vajna, S.} \emph{et~al.}
\newblock \bibinfo{journal}{\bibinfo{title}{Higher-order contributions to the
  rashba-bychkov effect with application to the bi/ag(111) surface alloy}}.
\newblock {\emph{\JournalTitle{Phys. Rev. B}}} \textbf{\bibinfo{volume}{85}},
  \bibinfo{pages}{075404} (\bibinfo{year}{2012}).
\newblock \urlprefix\url{http://link.aps.org/doi/10.1103/PhysRevB.85.075404}.
\newblock \doiprefix 10.1103/PhysRevB.85.075404.

\bibitem{NatComm_PST}
\bibinfo{author}{Tao, L.~L.} \& \bibinfo{author}{Tsymbal, E.~Y.}
\newblock \bibinfo{journal}{\bibinfo{title}{Persistent spin texture enforced by
  symmetry}}.
\newblock {\emph{\JournalTitle{Nature Communications}}}
  \textbf{\bibinfo{volume}{9}}, \bibinfo{pages}{2763} (\bibinfo{year}{2018}).
\newblock \urlprefix\url{https://doi.org/10.1038/s41467-018-05137-0}.
\newblock \doiprefix 10.1038/s41467-018-05137-0.

\bibitem{PhysRevLett.107.156803}
\bibinfo{author}{Park, S.~R.}, \bibinfo{author}{Kim, C.~H.},
  \bibinfo{author}{Yu, J.}, \bibinfo{author}{Han, J.~H.} \&
  \bibinfo{author}{Kim, C.}
\newblock \bibinfo{journal}{\bibinfo{title}{Orbital-angular-momentum based
  origin of rashba-type surface band splitting}}.
\newblock {\emph{\JournalTitle{Phys. Rev. Lett.}}}
  \textbf{\bibinfo{volume}{107}}, \bibinfo{pages}{156803}
  (\bibinfo{year}{2011}).
\newblock
  \urlprefix\url{https://link.aps.org/doi/10.1103/PhysRevLett.107.156803}.
\newblock \doiprefix 10.1103/PhysRevLett.107.156803.

\bibitem{CarlosMera_PRL}
\bibinfo{author}{{Mera Acosta}, C.} \& \bibinfo{author}{{Fazzio}, A.}
\newblock \bibinfo{journal}{\bibinfo{title}{{Spin-polarization control driven
  by a Rashba-type effect breaking the mirror symmetry in two-dimensional dual
  topological insulators}}}.
\newblock {\emph{\JournalTitle{ArXiv e-prints}}}  (\bibinfo{year}{2018}).
\newblock \eprint{1811.11014}.

\bibitem{Zeeman_type}
\bibinfo{author}{Yuan, H.} \emph{et~al.}
\newblock \bibinfo{journal}{\bibinfo{title}{Zeeman-type spin splitting
  controlled by an electric field}}.
\newblock {\emph{\JournalTitle{Nature Physics}}} \textbf{\bibinfo{volume}{9}},
  \bibinfo{pages}{563 EP --} (\bibinfo{year}{2013}).
\newblock \urlprefix\url{http://dx.doi.org/10.1038/nphys2691}.

\bibitem{PhysRevLett.108.196802}
\bibinfo{author}{Xiao, D.}, \bibinfo{author}{Liu, G.-B.},
  \bibinfo{author}{Feng, W.}, \bibinfo{author}{Xu, X.} \& \bibinfo{author}{Yao,
  W.}
\newblock \bibinfo{journal}{\bibinfo{title}{Coupled spin and valley physics in
  monolayers of ${\mathrm{mos}}_{2}$ and other group-vi dichalcogenides}}.
\newblock {\emph{\JournalTitle{Phys. Rev. Lett.}}}
  \textbf{\bibinfo{volume}{108}}, \bibinfo{pages}{196802}
  (\bibinfo{year}{2012}).
\newblock
  \urlprefix\url{https://link.aps.org/doi/10.1103/PhysRevLett.108.196802}.
\newblock \doiprefix 10.1103/PhysRevLett.108.196802.

\bibitem{PhysRevB.91.235204}
\bibinfo{author}{Liu, Q.} \emph{et~al.}
\newblock \bibinfo{journal}{\bibinfo{title}{Search and design of nonmagnetic
  centrosymmetric layered crystals with large local spin polarization}}.
\newblock {\emph{\JournalTitle{Phys. Rev. B}}} \textbf{\bibinfo{volume}{91}},
  \bibinfo{pages}{235204} (\bibinfo{year}{2015}).
\newblock \urlprefix\url{https://link.aps.org/doi/10.1103/PhysRevB.91.235204}.
\newblock \doiprefix 10.1103/PhysRevB.91.235204.

\bibitem{CURTAROLO2012227}
\bibinfo{author}{Curtarolo, S.} \emph{et~al.}
\newblock \bibinfo{journal}{\bibinfo{title}{Aflowlib.org: A distributed
  materials properties repository from high-throughput ab initio
  calculations}}.
\newblock {\emph{\JournalTitle{Computational Materials Science}}}
  \textbf{\bibinfo{volume}{58}}, \bibinfo{pages}{227 -- 235}
  (\bibinfo{year}{2012}).
\newblock
  \urlprefix\url{http://www.sciencedirect.com/science/article/pii/S0927025612000687}.
\newblock \doiprefix https://doi.org/10.1016/j.commatsci.2012.02.002.

\bibitem{TAYLOR2014178}
\bibinfo{author}{Taylor, R.~H.} \emph{et~al.}
\newblock \bibinfo{journal}{\bibinfo{title}{A restful api for exchanging
  materials data in the aflowlib.org consortium}}.
\newblock {\emph{\JournalTitle{Computational Materials Science}}}
  \textbf{\bibinfo{volume}{93}}, \bibinfo{pages}{178 -- 192}
  (\bibinfo{year}{2014}).
\newblock
  \urlprefix\url{http://www.sciencedirect.com/science/article/pii/S0927025614003322}.
\newblock \doiprefix https://doi.org/10.1016/j.commatsci.2014.05.014.

\bibitem{Goldschmidt1927}
\bibinfo{author}{Goldschmidt, V.}
\newblock \bibinfo{journal}{\bibinfo{title}{Geochemische verteilungsgesetze
  viii. bau und eigenschaften von krystallen}}.
\newblock {\emph{\JournalTitle{Skrifter utgitt av det Norske Videnskaps-Akademi
  i Oslo 1: Matematisk-Naturvidenskapelig Klasse}}}
  \textbf{\bibinfo{volume}{1927}}, \bibinfo{pages}{1--156}
  (\bibinfo{year}{1927}).

\bibitem{Jain2013}
\bibinfo{author}{Jain, A.} \emph{et~al.}
\newblock \bibinfo{journal}{\bibinfo{title}{{The Materials Project: A materials
  genome approach to accelerating materials innovation}}}.
\newblock {\emph{\JournalTitle{APL Materials}}} \textbf{\bibinfo{volume}{1}},
  \bibinfo{pages}{011002} (\bibinfo{year}{2013}).
\newblock
  \urlprefix\url{http://link.aip.org/link/AMPADS/v1/i1/p011002/s1\&Agg=doi}.
\newblock \doiprefix 10.1063/1.4812323.

\bibitem{Jain2011a}
\bibinfo{author}{Jain, A.} \emph{et~al.}
\newblock \bibinfo{journal}{\bibinfo{title}{{Formation enthalpies by mixing GGA
  and GGA + U calculations}}}.
\newblock {\emph{\JournalTitle{Physical Review B}}}
  \textbf{\bibinfo{volume}{84}}, \bibinfo{pages}{045115}
  (\bibinfo{year}{2011}).
\newblock \urlprefix\url{http://link.aps.org/doi/10.1103/PhysRevB.84.045115}.
\newblock \doiprefix 10.1103/PhysRevB.84.045115.

\bibitem{CHARIFI20091632}
\bibinfo{author}{Charifi, Z.}, \bibinfo{author}{Baaziz, H.} \&
  \bibinfo{author}{Hamad, B.}
\newblock \bibinfo{journal}{\bibinfo{title}{Theoretical prediction of the
  structural and electronic properties of pseudocubic x3as4 (x=c, si, ge and
  sn) compounds}}.
\newblock {\emph{\JournalTitle{Physica B: Condensed Matter}}}
  \textbf{\bibinfo{volume}{404}}, \bibinfo{pages}{1632 -- 1637}
  (\bibinfo{year}{2009}).
\newblock
  \urlprefix\url{http://www.sciencedirect.com/science/article/pii/S0921452609000453}.
\newblock \doiprefix https://doi.org/10.1016/j.physb.2009.01.043.

\bibitem{Badachhape1962}
\bibinfo{author}{Badachhape, S.} \& \bibinfo{author}{Goswami, A.}
\newblock \bibinfo{journal}{\bibinfo{title}{Structure of evaporated tin
  sulphide}}.
\newblock {\emph{\JournalTitle{Journal of the Physical Society of Japan,
  Supplement B2}}} \textbf{\bibinfo{volume}{17}}, \bibinfo{pages}{251--253}
  (\bibinfo{year}{1962}).

\bibitem{DuXiangP2010}
\bibinfo{author}{Po, D.~X.} \& \bibinfo{author}{Xu, W.~Y.}
\newblock \bibinfo{journal}{\bibinfo{title}{Investigation of osmium carbides
  with various stoichiometries: first-principles calculations}}.
\newblock {\emph{\JournalTitle{Journal of Applied Physics}}}
  \textbf{\bibinfo{volume}{107}}, \bibinfo{pages}{053506--1--053506--6}
  (\bibinfo{year}{2010}).

\bibitem{2127}
\bibinfo{author}{Luo, H.} \emph{et~al.}
\newblock \bibinfo{journal}{\bibinfo{title}{Origin of the z-28 rule in
  mn2cu-based heusler alloys: A comparing study}}.
\newblock {\emph{\JournalTitle{Journal of Magnetism and Magnetic Materials}}}
  \textbf{\bibinfo{volume}{324}}, \bibinfo{pages}{2127 -- 2130}
  (\bibinfo{year}{2012}).
\newblock
  \urlprefix\url{http://www.sciencedirect.com/science/article/pii/S0304885312001230}.
\newblock \doiprefix https://doi.org/10.1016/j.jmmm.2012.02.026.

\bibitem{Raub1962}
\bibinfo{author}{Raub, E.} \& \bibinfo{author}{Fritzsche, W.}
\newblock \bibinfo{journal}{\bibinfo{title}{Germanium-ruthenium legierungen}}.
\newblock {\emph{\JournalTitle{Zeitschrift fuer Metallkunde}}}
  \textbf{\bibinfo{volume}{53}}, \bibinfo{pages}{779--781}
  (\bibinfo{year}{1962}).

\bibitem{Goeransson1995}
\bibinfo{author}{Goeransson, K.}, \bibinfo{author}{Engstroem, I.} \&
  \bibinfo{author}{Nolaeng, B.}
\newblock \bibinfo{journal}{\bibinfo{title}{Structure refinements for some
  platinum metal monosilicides}}.
\newblock {\emph{\JournalTitle{Journal of Alloys and Compounds}}}
  \textbf{\bibinfo{volume}{219}}, \bibinfo{pages}{107--110}
  (\bibinfo{year}{1995}).

\bibitem{Vocadlo2002}
\bibinfo{author}{Vocadlo, L.}, \bibinfo{author}{Knight, K.},
  \bibinfo{author}{Price, G.} \& \bibinfo{author}{Wood, I.}
\newblock \bibinfo{journal}{\bibinfo{title}{Thermal expansion and crystal
  structure of fesi between 4 and 1173 k}}.
\newblock {\emph{\JournalTitle{Physics and Chemistry of Minerals}}}
  \textbf{\bibinfo{volume}{29}}, \bibinfo{pages}{132--139}
  (\bibinfo{year}{2002}).

\bibitem{Wood1996}
\bibinfo{author}{Wood, I.}, \bibinfo{author}{David, W.}, \bibinfo{author}{Hull,
  S.} \& \bibinfo{author}{Price, G.}
\newblock \bibinfo{journal}{\bibinfo{title}{A high-pressure study of
  epsilon-(fe si), between 0 and 8.5 gpa, by time-of-flight neutron powder
  diffraction}}.
\newblock {\emph{\JournalTitle{Journal of Applied Crystallography}}}
  \textbf{\bibinfo{volume}{29}}, \bibinfo{pages}{215--218}
  (\bibinfo{year}{1996}).

\bibitem{Pauling1948}
\bibinfo{author}{Pauling, L.} \& \bibinfo{author}{Soldate, A.}
\newblock \bibinfo{journal}{\bibinfo{title}{The nature of the bonds in the iron
  silicide fesi and related crystals}}.
\newblock {\emph{\JournalTitle{Acta Crystallographica (1,1948-23,1967)}}}
  \textbf{\bibinfo{volume}{1}}, \bibinfo{pages}{212--216}
  (\bibinfo{year}{1948}).

\bibitem{Weitzer1988}
\bibinfo{author}{Weitzer, F.}, \bibinfo{author}{Rogl, P.} \&
  \bibinfo{author}{Schuster, J.}
\newblock \bibinfo{journal}{\bibinfo{title}{X-ray investigations in the systems
  ruthenium-silicon and ruthenium-silicon-nitrogen}}.
\newblock {\emph{\JournalTitle{Zeitschrift fuer Metallkunde}}}
  \textbf{\bibinfo{volume}{79}}, \bibinfo{pages}{154--156}
  (\bibinfo{year}{1988}).

\bibitem{Weitzer1997}
\bibinfo{author}{Weitzer, F.}, \bibinfo{author}{Perring, L.},
  \bibinfo{author}{Gachon, J.}, \bibinfo{author}{Feschotte, P.} \&
  \bibinfo{author}{Schuster, J.}
\newblock \bibinfo{journal}{\bibinfo{title}{Structure refinements of some
  compounds of the ru-si, ru-ge and ru-sn systems}}.
\newblock {\emph{\JournalTitle{Electrochemical Society Proceedings}}}
  \textbf{\bibinfo{volume}{39}}, \bibinfo{pages}{241--249}
  (\bibinfo{year}{1997}).

\bibitem{Finnie1962}
\bibinfo{author}{Finnie, L.}
\newblock \bibinfo{journal}{\bibinfo{title}{Structures and compositions of the
  silicides of ruthenium, osmium, rhodium, and iridium}}.
\newblock {\emph{\JournalTitle{Journal of the Less-Common Metals}}}
  \textbf{\bibinfo{volume}{4}}, \bibinfo{pages}{24--34} (\bibinfo{year}{1962}).

\bibitem{Mattheiss1992}
\bibinfo{author}{Mattheiss, L.}
\newblock \bibinfo{journal}{\bibinfo{title}{Calculated structural properties of
  crsi2,mosi2 and wsi2}}.
\newblock {\emph{\JournalTitle{Physical Review, Serie 3. B - Condensed Matter
  (18,1978-)}}} \textbf{\bibinfo{volume}{45}}, \bibinfo{pages}{3252--3259}
  (\bibinfo{year}{1992}).

\bibitem{Heurle1980}
\bibinfo{author}{d'Heurle, F.}, \bibinfo{author}{Petersson, C.} \&
  \bibinfo{author}{Tsai, M.}
\newblock \bibinfo{journal}{\bibinfo{title}{Observations on the hexagonal form
  of mo si2 and w si2 films produced by ion implantation and on related
  snowplow effects}}.
\newblock {\emph{\JournalTitle{Journal of Applied Physics}}}
  \textbf{\bibinfo{volume}{51}}, \bibinfo{pages}{5976--5980}
  (\bibinfo{year}{1980}).

\bibitem{SOTO20121}
\bibinfo{author}{Soto, G.}
\newblock \bibinfo{journal}{\bibinfo{title}{Computational study of hf, ta, w,
  re, ir, os and pt pernitrides}}.
\newblock {\emph{\JournalTitle{Computational Materials Science}}}
  \textbf{\bibinfo{volume}{61}}, \bibinfo{pages}{1 -- 5}
  (\bibinfo{year}{2012}).
\newblock
  \urlprefix\url{http://www.sciencedirect.com/science/article/pii/S0927025612002030}.
\newblock \doiprefix https://doi.org/10.1016/j.commatsci.2012.03.056.

\bibitem{Matsumoto2010}
\bibinfo{author}{Matsumoto, A.}, \bibinfo{author}{Koyama, Y.} \&
  \bibinfo{author}{Tanaka, I.}
\newblock \bibinfo{journal}{\bibinfo{title}{Structures and energetics of bi2 o3
  polymorphs in a defective fluorite family derived by systematic
  first-principles lattice dynamics calculations}}.
\newblock {\emph{\JournalTitle{Physical Review, Serie 3. B - Condensed Matter
  (18,1978-)}}} \textbf{\bibinfo{volume}{81}},
  \bibinfo{pages}{094117--1--094117--11} (\bibinfo{year}{2010}).

\bibitem{Jellinek1960}
\bibinfo{author}{Jellinek, F.}, \bibinfo{author}{Brauer, G.} \&
  \bibinfo{author}{Mueller, H.}
\newblock \bibinfo{journal}{\bibinfo{title}{Molybdenum and niobium sulphides}}.
\newblock {\emph{\JournalTitle{Nature (London)}}}
  \textbf{\bibinfo{volume}{185}}, \bibinfo{pages}{376--377}
  (\bibinfo{year}{1960}).

\bibitem{Traill1962}
\bibinfo{author}{Traill, R.}
\newblock \bibinfo{journal}{\bibinfo{title}{A rhombohedral polytype of
  molybdenite}}.
\newblock {\emph{\JournalTitle{Canadian Mineralogist}}}
  \textbf{\bibinfo{volume}{7}}, \bibinfo{pages}{524--526}
  (\bibinfo{year}{1962}).

\bibitem{Takeuchi1964}
\bibinfo{author}{Takeuchi, Y.} \& \bibinfo{author}{Nowacki, W.}
\newblock \bibinfo{journal}{\bibinfo{title}{Detailed crystal structure of
  rhombohedral mo s2 and systematic deduction of possible polytypes of
  molybdenite}}.
\newblock {\emph{\JournalTitle{Schweizerische Mineralogische und
  Petrographische Mitteilungen}}} \textbf{\bibinfo{volume}{44}},
  \bibinfo{pages}{105--120} (\bibinfo{year}{1964}).

\bibitem{Schutte1987}
\bibinfo{author}{Schutte, W.}, \bibinfo{author}{de~Boer, J.} \&
  \bibinfo{author}{Jellinek, F.}
\newblock \bibinfo{journal}{\bibinfo{title}{Crystal structures of tungsten
  disulfide and diselenide}}.
\newblock {\emph{\JournalTitle{Journal of Solid State Chemistry}}}
  \textbf{\bibinfo{volume}{70}}, \bibinfo{pages}{207--209}
  (\bibinfo{year}{1987}).

\bibitem{Wildervanck1964}
\bibinfo{author}{Wildervanck, J.} \& \bibinfo{author}{Jellinek, F.}
\newblock \bibinfo{journal}{\bibinfo{title}{Preparation and crystallinity of
  molybdenum and tungsten sulfides}}.
\newblock {\emph{\JournalTitle{Zeitschrift fuer Anorganische und Allgemeine
  Chemie (1950) (DE)}}} \textbf{\bibinfo{volume}{328}},
  \bibinfo{pages}{309--318} (\bibinfo{year}{1964}).

\bibitem{Flieher1968}
\bibinfo{author}{Flieher, G.}, \bibinfo{author}{Voellenkle, H.} \&
  \bibinfo{author}{Nowotny, H.}
\newblock \bibinfo{journal}{\bibinfo{title}{Die kristallstruktur von ir4 ge5}}.
\newblock {\emph{\JournalTitle{Monatshefte fuer Chemie}}}
  \textbf{\bibinfo{volume}{99}}, \bibinfo{pages}{877--883}
  (\bibinfo{year}{1968}).

\bibitem{Bhan1960}
\bibinfo{author}{Bhan, S.} \& \bibinfo{author}{Schubert, K.}
\newblock \bibinfo{journal}{\bibinfo{title}{Zum aufbau der systeme kobalt -
  germanium, rhodium - silizium sowie einiger verwandter legierungen}}.
\newblock {\emph{\JournalTitle{Zeitschrift fuer Metallkunde}}}
  \textbf{\bibinfo{volume}{51}}, \bibinfo{pages}{327--339}
  (\bibinfo{year}{1960}).

\bibitem{Panday1967}
\bibinfo{author}{Panday, P.}, \bibinfo{author}{Singh, G.} \&
  \bibinfo{author}{Schubert, K.}
\newblock \bibinfo{journal}{\bibinfo{title}{Kristallstruktur von ir4 ge5}}.
\newblock {\emph{\JournalTitle{Zeitschrift fuer Kristallographie,
  Kristallgeometrie, Kristallphysik, Kristallchemie (-144,1977)}}}
  \textbf{\bibinfo{volume}{125}}, \bibinfo{pages}{274--285}
  (\bibinfo{year}{1967}).

\bibitem{Bhan1970}
\bibinfo{author}{Bhan, S.} \& \bibinfo{author}{Schubert, K.}
\newblock \bibinfo{journal}{\bibinfo{title}{Kristallstruktur von tl5 te3 und
  tl2 te3}}.
\newblock {\emph{\JournalTitle{Journal of the Less-Common Metals}}}
  \textbf{\bibinfo{volume}{20}}, \bibinfo{pages}{229--235}
  (\bibinfo{year}{1970}).

\bibitem{Uno1988}
\bibinfo{author}{Uno, R.} \emph{et~al.}
\newblock \bibinfo{journal}{\bibinfo{title}{Powder diffractometry at the
  tsukuba photon factory}}.
\newblock {\emph{\JournalTitle{Australian Journal of Physics}}}
  \textbf{\bibinfo{volume}{41}}, \bibinfo{pages}{133--144}
  (\bibinfo{year}{1988}).
\newblock \urlprefix\url{https://doi.org/10.1071/PH880133}.

\bibitem{Verkelis1974}
\bibinfo{author}{Verkelis, J.}
\newblock \bibinfo{journal}{\bibinfo{title}{Single crystals of in2 te5}}.
\newblock {\emph{\JournalTitle{Materials Research Bulletin}}}
  \textbf{\bibinfo{volume}{9}}, \bibinfo{pages}{1063--1065}
  (\bibinfo{year}{1974}).

\bibitem{Sutherland1976}
\bibinfo{author}{Sutherland, H.}, \bibinfo{author}{Hogg, J.} \&
  \bibinfo{author}{Walton, P.}
\newblock \bibinfo{journal}{\bibinfo{title}{Indium polytelluride in2 te5}}.
\newblock {\emph{\JournalTitle{Acta Crystallographica B (24,1968-38,1982)}}}
  \textbf{\bibinfo{volume}{32}}, \bibinfo{pages}{2539--2541}
  (\bibinfo{year}{1976}).

\bibitem{Boettcher1980}
\bibinfo{author}{Boettcher, P.}
\newblock \bibinfo{journal}{\bibinfo{title}{Synthesis and crystal structure of
  rb2 te3 and cs2 te3}}.
\newblock {\emph{\JournalTitle{Journal of the Less-Common Metals}}}
  \textbf{\bibinfo{volume}{70}}, \bibinfo{pages}{263--271}
  (\bibinfo{year}{1980}).

\bibitem{Chuntonov1982}
\bibinfo{author}{Chuntonov, K.}, \bibinfo{author}{Orlov, A.},
  \bibinfo{author}{Yatsenko, S.}, \bibinfo{author}{Grin', Y.} \&
  \bibinfo{author}{Miroshnikova, L.}
\newblock \bibinfo{journal}{\bibinfo{title}{Synthesis of chalcogenides with the
  composition a2(i) b3(vi) and crystal structure of rb2 te3 and cs2 te3}}.
\newblock {\emph{\JournalTitle{Izvestiya Akademii Nauk SSSR, Neorganicheskie
  Materialy}}} \textbf{\bibinfo{volume}{18}}, \bibinfo{pages}{1113--1116}
  (\bibinfo{year}{1982}).

\bibitem{Alex_NatChem}
\bibinfo{author}{Zunger, A.}
\newblock \bibinfo{journal}{\bibinfo{title}{Inverse design in search of
  materials with target functionalities}}.
\newblock {\emph{\JournalTitle{Nature Reviews Chemistry}}}
  \textbf{\bibinfo{volume}{2}}, \bibinfo{pages}{0121 EP --}
  (\bibinfo{year}{2018}).
\newblock \urlprefix\url{http://dx.doi.org/10.1038/s41570-018-0121}.

\bibitem{PhysRevLett.102.056405}
\bibinfo{author}{Luo, J.-W.}, \bibinfo{author}{Bester, G.} \&
  \bibinfo{author}{Zunger, A.}
\newblock \bibinfo{journal}{\bibinfo{title}{Full-zone spin splitting for
  electrons and holes in bulk gaas and gasb}}.
\newblock {\emph{\JournalTitle{Phys. Rev. Lett.}}}
  \textbf{\bibinfo{volume}{102}}, \bibinfo{pages}{056405}
  (\bibinfo{year}{2009}).
\newblock
  \urlprefix\url{https://link.aps.org/doi/10.1103/PhysRevLett.102.056405}.
\newblock \doiprefix 10.1103/PhysRevLett.102.056405.

\bibitem{PhysRevB.50.2715}
\bibinfo{author}{Yeh, C.-Y.}, \bibinfo{author}{Wei, S.-H.} \&
  \bibinfo{author}{Zunger, A.}
\newblock \bibinfo{journal}{\bibinfo{title}{Relationships between the band gaps
  of the zinc-blende and wurtzite modifications of semiconductors}}.
\newblock {\emph{\JournalTitle{Phys. Rev. B}}} \textbf{\bibinfo{volume}{50}},
  \bibinfo{pages}{2715--2718} (\bibinfo{year}{1994}).
\newblock \urlprefix\url{https://link.aps.org/doi/10.1103/PhysRevB.50.2715}.
\newblock \doiprefix 10.1103/PhysRevB.50.2715.

\bibitem{PhysRevLett.121.077701}
\bibinfo{author}{Marcellina, E.} \emph{et~al.}
\newblock \bibinfo{journal}{\bibinfo{title}{Electrical control of the zeeman
  spin splitting in two-dimensional hole systems}}.
\newblock {\emph{\JournalTitle{Phys. Rev. Lett.}}}
  \textbf{\bibinfo{volume}{121}}, \bibinfo{pages}{077701}
  (\bibinfo{year}{2018}).
\newblock
  \urlprefix\url{https://link.aps.org/doi/10.1103/PhysRevLett.121.077701}.
\newblock \doiprefix 10.1103/PhysRevLett.121.077701.

\bibitem{PhysRevLett.77.3865}
\bibinfo{author}{Perdew, J.~P.}, \bibinfo{author}{Burke, K.} \&
  \bibinfo{author}{Ernzerhof, M.}
\newblock \bibinfo{journal}{\bibinfo{title}{Generalized gradient approximation
  made simple}}.
\newblock {\emph{\JournalTitle{Phys. Rev. Lett.}}}
  \textbf{\bibinfo{volume}{77}}, \bibinfo{pages}{3865--3868}
  (\bibinfo{year}{1996}).
\newblock \urlprefix\url{http://link.aps.org/doi/10.1103/PhysRevLett.77.3865}.
\newblock \doiprefix 10.1103/PhysRevLett.77.3865.

\bibitem{PhysRevB.52.R5467}
\bibinfo{author}{Liechtenstein, A.~I.}, \bibinfo{author}{Anisimov, V.~I.} \&
  \bibinfo{author}{Zaanen, J.}
\newblock \bibinfo{journal}{\bibinfo{title}{Density-functional theory and
  strong interactions: Orbital ordering in mott-hubbard insulators}}.
\newblock {\emph{\JournalTitle{Phys. Rev. B}}} \textbf{\bibinfo{volume}{52}},
  \bibinfo{pages}{R5467--R5470} (\bibinfo{year}{1995}).
\newblock \urlprefix\url{https://link.aps.org/doi/10.1103/PhysRevB.52.R5467}.
\newblock \doiprefix 10.1103/PhysRevB.52.R5467.

\bibitem{PhysRevB.57.1505}
\bibinfo{author}{Dudarev, S.~L.}, \bibinfo{author}{Botton, G.~A.},
  \bibinfo{author}{Savrasov, S.~Y.}, \bibinfo{author}{Humphreys, C.~J.} \&
  \bibinfo{author}{Sutton, A.~P.}
\newblock \bibinfo{journal}{\bibinfo{title}{Electron-energy-loss spectra and
  the structural stability of nickel oxide: An lsda+u study}}.
\newblock {\emph{\JournalTitle{Phys. Rev. B}}} \textbf{\bibinfo{volume}{57}},
  \bibinfo{pages}{1505--1509} (\bibinfo{year}{1998}).
\newblock \urlprefix\url{https://link.aps.org/doi/10.1103/PhysRevB.57.1505}.
\newblock \doiprefix 10.1103/PhysRevB.57.1505.

\bibitem{PhysRevB.54.11169}
\bibinfo{author}{Kresse, G.} \& \bibinfo{author}{Furthm\"uller, J.}
\newblock \bibinfo{journal}{\bibinfo{title}{Efficient iterative schemes for
  \textit{ab initio} total-energy calculations using a plane-wave basis set}}.
\newblock {\emph{\JournalTitle{Phys. Rev. B}}} \textbf{\bibinfo{volume}{54}},
  \bibinfo{pages}{11169--11186} (\bibinfo{year}{1996}).
\newblock \urlprefix\url{http://link.aps.org/doi/10.1103/PhysRevB.54.11169}.
\newblock \doiprefix 10.1103/PhysRevB.54.11169.

\bibitem{PhysRevB.59.1758}
\bibinfo{author}{Kresse, G.} \& \bibinfo{author}{Joubert, D.}
\newblock \bibinfo{journal}{\bibinfo{title}{From ultrasoft pseudopotentials to
  the projector augmented-wave method}}.
\newblock {\emph{\JournalTitle{Phys. Rev. B}}} \textbf{\bibinfo{volume}{59}},
  \bibinfo{pages}{1758--1775} (\bibinfo{year}{1999}).
\newblock \urlprefix\url{https://link.aps.org/doi/10.1103/PhysRevB.59.1758}.
\newblock \doiprefix 10.1103/PhysRevB.59.1758.

\end{thebibliography}

\section*{Acknowledgements}
This work was financed by FAPESP (grants 18/11856-7 and 17/02317-2), and CNPq. This work was performed using the computational infrastructure of the LNCC supercomputer center (Santos Dumont).

\end{document}